\documentclass[twocolumn,english,aps,superscriptaddress,amsmath,amssymb,floatfix]{revtex4-1}
\usepackage[colorlinks=true,citecolor=blue,linkcolor=magenta]{hyperref}

\usepackage[markup=blue, authormarkupposition=left]{changes} 

\usepackage{soul}
\usepackage[utf8]{inputenc}
\usepackage[english]{babel}
\usepackage{amsmath,amsfonts,amssymb}
\usepackage[T1]{fontenc}
\usepackage{url}

\usepackage{cancel}
\usepackage{amsmath}
\usepackage{amsfonts}
\usepackage{amssymb}

\usepackage{epstopdf}
\usepackage{graphicx}

\begin{document}

\title{Photonic Millimeter-wave Generation Beyond the Cavity Thermal Limit}

%\title{Record Low Noise Millimeter-Wave Generation via \\Photonic SIL Lasers and Ultrastable Fabry-Perot Heterodyne}

\author{William Groman}
\affiliation{Electrical Computer \& Energy Engineering, University of Colorado, Boulder, CO 80309, USA}
\affiliation{Department of Physics, University of Colorado Boulder, 440 UCB Boulder, CO 80309, USA}
\affiliation{National Institute of Standards and Technology, 325 Broadway, Boulder, CO 80305, USA}

\author{Igor Kudelin}
\affiliation{Electrical Computer \& Energy Engineering, University of Colorado, Boulder, CO 80309, USA}
\affiliation{Department of Physics, University of Colorado Boulder, 440 UCB Boulder, CO 80309, USA}
\affiliation{National Institute of Standards and Technology, 325 Broadway, Boulder, CO 80305, USA}

\author{Alexander Lind}
\affiliation{Electrical Computer \& Energy Engineering, University of Colorado, Boulder, CO 80309, USA}
\affiliation{National Institute of Standards and Technology, 325 Broadway, Boulder, CO 80305, USA}

\author{Dahyeon Lee}
\affiliation{Department of Physics, University of Colorado Boulder, 440 UCB Boulder, CO 80309, USA}
\affiliation{National Institute of Standards and Technology, 325 Broadway, Boulder, CO 80305, USA}

\author{Takuma Nakamura}
\affiliation{National Institute of Standards and Technology, 325 Broadway, Boulder, CO 80305, USA}

\author{Yifan Liu}
\affiliation{Department of Physics, University of Colorado Boulder, 440 UCB Boulder, CO 80309, USA}
\affiliation{National Institute of Standards and Technology, 325 Broadway, Boulder, CO 80305, USA}

\author{Megan L. Kelleher}
\affiliation{Department of Physics, University of Colorado Boulder, 440 UCB Boulder, CO 80309, USA}
\affiliation{National Institute of Standards and Technology, 325 Broadway, Boulder, CO 80305, USA}

\author{Charles A. McLemore}
\affiliation{Department of Physics, University of Colorado Boulder, 440 UCB Boulder, CO 80309, USA}
\affiliation{National Institute of Standards and Technology, 325 Broadway, Boulder, CO 80305, USA}

\author{Joel Guo}
\affiliation{Department of Electrical and Computer Engineering, University of California, Santa Barbara, Santa Barbara, CA 93106, USA}

\author{Lue Wu}
\affiliation{Department of Electrical and Computer Engineering, University of California, Santa Barbara, Santa Barbara, CA 93106, USA}

\author{Warren Jin}
\affiliation{Department of Electrical and Computer Engineering, University of California, Santa Barbara, Santa Barbara, CA 93106, USA}

\author{John E. Bowers}
\affiliation{Department of Electrical and Computer Engineering, University of California, Santa Barbara, Santa Barbara, CA 93106, USA}

\author{Franklyn Quinlan}
\affiliation{Electrical Computer \& Energy Engineering, University of Colorado, Boulder, CO 80309, USA}
\affiliation{National Institute of Standards and Technology, 325 Broadway, Boulder, CO 80305, USA}

\author{Scott A. Diddams}
\affiliation{Electrical Computer \& Energy Engineering, University of Colorado, Boulder, CO 80309, USA}
\affiliation{Department of Physics, University of Colorado Boulder, 440 UCB Boulder, CO 80309, USA}
\affiliation{National Institute of Standards and Technology, 325 Broadway, Boulder, CO 80305, USA}

\begin{abstract}

Next-generation communications, radar and navigation systems will extend and exploit the higher bandwidth of the millimeter-wave domain for increased communication data rates as well as radar with higher sensitivity and increased spatial resolution. However, realizing these advantages will require the generation of millimeter-wave signals with low phase noise in simple and compact form-factors. The rapidly developing field of photonic integration addresses this challenge and provides a path toward simplified and portable, low-noise mm-wave generation for these applications. We leverage these advances by heterodyning two silicon photonic chip lasers, phase-locked to the same miniature Fabry-Perot (F-P) cavity to demonstrate a simple framework for generating low-noise millimeter-waves with phase noise below the thermal limit of the F-P cavity. Specifically, we generate 94.5 GHz and 118.1 GHz millimeter-wave signals with phase noise of $-117$~dBc/Hz at 10 kHz offset, decreasing to $-120$ dBc/Hz at 40 kHz offset, a record low value for such photonic devices. We achieve this with existing  technologies that can be integrated into a platform less than $\approx 10 \hspace{3 pt}\text{mL}$ in volume. Our work illustrates the significant potential and advantages of low size, weight, and power (SWaP) photonic-sourced mm-waves for communications and sensing. 

\end{abstract}

\maketitle

\section*{Introduction}

Inherent to the higher frequency of millimeter-wave carriers (mm-wave, $\approx 30\to 300$ GHz) is the capability of higher modulation bandwidth and faster data transfer rates \cite{Ma2018, Akyildiz2014}, finer radar and sensing spatial resolution \cite{Soumya2023, Rappaport2019}, antenna miniaturization \cite{Soumya2023}, and viability of atmospheric transmission \cite{Rappaport2013}. The impact of mm-waves is already evident in cutting-edge telecommunications, with 5G utilizing carriers in both the microwave and low mm-wave range ($\lessapprox 50 \hspace{3 pt}\text{GHz}$), and 6G technology pushing carrier frequencies beyond $90 \hspace{3 pt}\text{GHz}$ \cite{Hong2021}. 
Furthermore, mm-wave technologies have already shown promise for safe and non-invasive biomedical and security sensing applications \cite{Wang2023, Luukanen2012}, and are widely used as radio astronomy references \cite{doeleman2009}.

%Precise Positioning, Imaging, Scanning, 1 Pbps possible data rates

In the microwave range, rack-mount-volume synthesizers dominate the application space. These provide wide tunability, but with close-to-carrier phase noise that, in the best case, is given by a frequency-multiplied quartz reference or other dielectric oscillator. Such signal sources are commonly extended to the mm-wave via carrier multiplication by $N$, but that comes with additional degradation in phase noise power spectral density by at least $20\log(N)$ \cite{Bara-Maillet2015}. As a result, commercial mm-wave synthesizers at 100 GHz are typically limited in phase-noise performance in the range of -100 dBc/Hz at 10 kHz offset and can struggle to meet the demands as local oscillators (LOs) in the aforementioned radar and telecommunication applications. On the other hand, laboratory research synthesizers provide remarkably low noise at the cost of increasing complexity and larger volume \cite{Bara2012}. 
%\cite{BerkeleyNucleonics, MiWave, Eravant, Spacek}%\cite{BerkeleyNucleonics, MiWave}
\begin{figure*}[t]
  \centering
  \includegraphics[width = \linewidth]{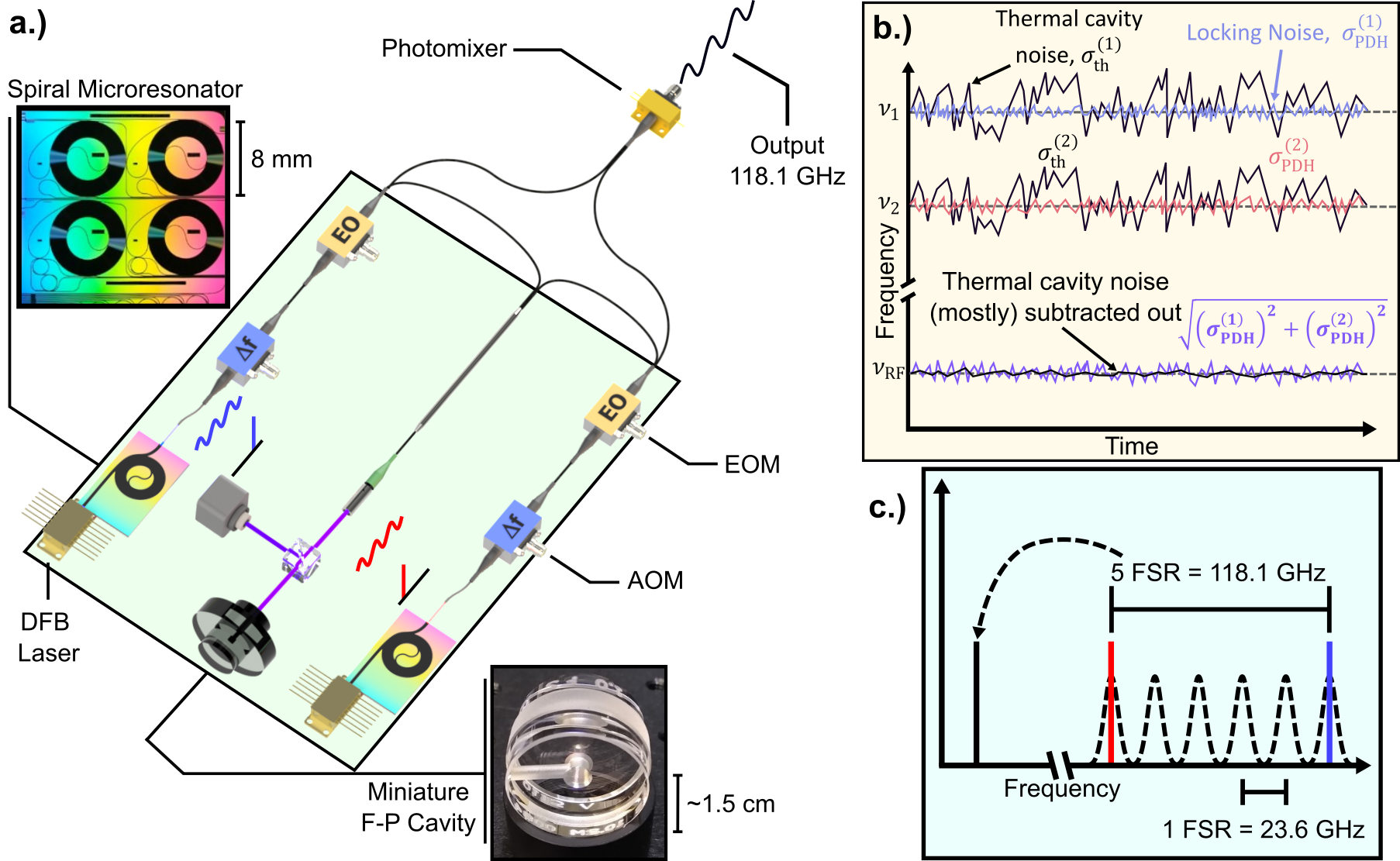}
\caption{a.) Simplified schematic of the experiment, with pictures of the key components zoomed in. b.) Illustration of the mechanism of common mode rejection. Two signals ($\nu_1$ and $\nu_2$) are indicated by their nominal frequencies (gray dashed lines) with their noise contributions (the thermal cavity noise of each laser, $\sigma^{(1)}_{\text{th}}$ and $\sigma^{(2)}_{\text{th}}$, respectively as well as PDH locking noise, $\sigma^{(1)}_{\text{PDH}}$ and $\sigma^{(2)}_{\text{PDH}}$, respectively) shown as jagged continuous lines. The signal that results from their heterodyning, $\nu_{\text{RF}}$, is shown with most of the thermal cavity noise being subtracted out. c.) Illustration of the two laser signals (red and blue) locked to two cavity modes (dashed Gaussians), and heterodyned at 5 cavity FSRs apart, to generate a 118.1 GHz RF tone. 
}
 \label{1}
 \vspace{-.32 cm}
\end{figure*} 

%SD version here.  Attempted to tighten up story of the first few paragraphs.  I saved Will's version below.  

Within this context, the field of mm-wave photonic synthesis with narrow linewidth lasers provides a compelling alternative to electronic approaches. Most simply, the interference of two continuous wave (CW) lasers can be photomixed (photodetected) to produce a mm-wave signal at the difference frequency of the lasers. This concept was first demonstrated just a few years after the invention of the laser, but progress has accelerated over the past two decades with multiple techniques involving lasers, electro-optics, and frequency combs to generate and control the relative frequency difference between the CW lasers \cite{Waltman1996,fukushima2003optoelectronic,Nagatsuma2009,rolland2014narrow,Fortier2016,Tetsumoto2021,Kittlaus2021}. 

Even greater opportunities for photonic mm-waves are offered by the confluence of recent advances in chip-integrated low-noise lasers and frequency combs and the means to frequency-stabilize these light sources with compact, high-finesse, and integrable high-Q optical cavities.  Importantly, lasers can now be  heterogeneously integrated on a CMOS-compatible platform together with extremely low-loss Si$_4$N$_3$ ring resonators for self-injection locking (SIL) operation \cite{Jin2021,Xiang2023}.  This leads to laser frequency noise that is typically associated with narrow-linewidth fiber lasers, but with smaller footprint \cite{Li2021}. In addition, miniaturization and integration of centimeter-sized optical cavities with finesse approaching $10^6$ \cite{Cheng2023, Liu2024} transforms the realization of hertz-linewidth lasers from large laboratory setups to the chip-scale, and in some cases even forgoing the need for vacuum enclosures \cite{Liu2024}. Recent combinations of such integrated photonic hardware have further utilized the powerful approach of optical frequency division (OFD) for low-noise microwave and mm-wave generation \cite{Kudelin2024,sun2024integrated}.

In this work, we show that the benefits of OFD are not required to further advance the phase noise of mm-wave generation at 100 GHz and above. Instead we exploit the high degree of thermal noise correlation between the axial modes of a Fabry-Perot (F-P) to strongly reject this noise in mm-wave generation. A cavity Q-factor approaching $10^{10}$ and linewidth near 20 kHz allows us to tightly lock the frequencies of two lasers to cavity modes with residual noise that is well below the room-temperature cavity thermal noise.  Thus, the correlated thermal noise of the cavity close-to-carrier that is imprinted on each laser is subtracted in the mm-wave heterodyne beat. 

We implement this concept with two SIL lasers that are frequency-locked to modes of a miniature high-Q Fabry-Perot cavity with free spectral range, $\text{FSR}=23.6$ GHz. For modes separated by  $5\times \text{FSR}=118$ GHz, the phase noise of the heterodyne beat is $-117.6 \hspace{3 pt}\text{dBc/Hz}$ at $10 \hspace{3pt}\text{kHz}$ offset, decreasing to $-124 \hspace{3 pt}\text{dBc/Hz}$ at $30 \hspace{3pt}\text{kHz}$. Importantly, our approach uses the same components that have been shown to be compatible with chip-scale integration, while offering phase noise levels that match those achieved with more complicated integrated OFD approaches. And at 10 kHz offset our results are close to the noise levels achieved in the very best photonic mm-wave laboratory experiments.  

Our straightforward approach to mm-wave generation effectively removes the fundamental limitations imposed by cavity thermal noise. As such, we anticipate that further improvements in phase-noise performance will be allowed to proceed at the more rapid pace of technical innovations in low-noise chip-integrated lasers and frequency control techniques.

\section*{Concept \& Experiment}

A simplified schematic of the experiment is shown in Fig.~\ref{1}a. Starting from the left, two distributed-feedback (DFB) lasers ($\nu_1$ and $\nu_2$) are individually coupled and self-injection locked to separate high~Q ($\text{Q} \gtrapprox 10^8$) Si$_3$N$_4$ microresonators, with free spectral ranges (FSRs) of about $135 \hspace{3 pt}\text{MHz}$. The self-injection locking is controlled by fine-tuning the facet distance (hence phase) of the DFBs relative to their respective microresonator chips, and results in a phase noise suppression of the free-running DFB signal by up to $50 \hspace{3 pt}\text{dB}$ \cite{Guo2022}. The SIL laser output is then fiber-coupled out of the chip for use. 

This initial self-injection locking suppresses the close-to-carrier noise and allows us to use the Pound-Drever-Hall (PDH) technique \cite{Drever1983} to subsequently lock each laser's frequency to an ultra-stable, temperature- and vacuum-controlled, miniature F-P cavity. In this case, the laser frequency inherits the stability of the cavity, but with frequency noise fundamentally limited by thermally driven stochastic fluctuations in the cavity length \cite{Numata2004}. Here, the cavity we used is exhibited and referred to in \cite{kelleher2023compact} as the "all-ULE cavity". %It has a Q $\gtrapprox 10^{10}$, and an FSR of $23.6 \hspace{3 pt}\text{GHz}$. 
We lock both lasers to separate F-P cavity modes such that the mm-wave carrier frequency is $f_{\mu}=\text{n}\times \text{FSR}=\nu_1-\nu_2$, as portrayed in Fig. \ref{1}c. Importantly, subtraction of these two PDH-locked signals, as occurs in photodetection, results in noise cancellation of the common noise, including thermal noise inherited by both lasers from the cavity. This yields an output signal with phase noise below that of the cavity thermal limit. 

Fig. \ref{1}b illustrates the primary noise contributions to $f_{\mu}$ and their subtraction in the time domain. The two PDH-locked lasers have frequency fluctuations dominated by two terms: the intrinsic cavity thermal noise $\sigma_{th}$ and the smaller residual noise of the PDH lock $\sigma_\text{PDH}$ that is related to technical servo limitations. Because of significant overlap of the spatial modes of $\nu_1$ and $\nu_2$ in the F-P cavity, there is strong amplitude and phase correlation between $\sigma_{th}^{(1)}$ and $\sigma_{th}^{(2)}$. In the heterodyne these noise terms largely subtract, as shown in the bottom of Fig. \ref{1}b.  However, the PDH lock noise $\sigma_\text{PDH}^{(1)}$ and $\sigma_\text{PDH}^{(2)}$ arises from uncorrelated shot noise from the separate laser locks and instead adds in quadrature in the heterodyne signal, such that $\sigma_{\mu}^2 = (\sigma^{(1)}_\text{PDH})^2+(\sigma^{(2)}_\text{PDH})^2$.

The extent of the rejection of the correlated (common-mode) cavity noise can be estimated by considering that the F-P cavity itself functions as a frequency divider. The magnitude of the thermally driven fluctuations of a frequency mode $\nu$ of the cavity is directly linked to the fluctuations of the cavity FSR, such that $\sigma_\text{FSR}=(\text{FSR}/\nu)\sigma_{\nu}$. Here we see that the frequency fluctuations $\sigma_{\nu}$ of mode $\nu$ are ideally reduced by the ratio of $\text{FSR}/\nu\approx 10^{-4}$. Similarly, $\sigma_{n\times\text{FSR}}=(n\times \text{FSR}/\nu)\sigma_{\nu}$, where $n\times\text{FSR}$ is the separation of any two modes of the cavity \cite{Liu2024}. Considering two lasers locked to modes separated by  $5\times\text{FSR}=118$ GHz, the expected common rejection of frequency fluctuations, is $6\times10^{-4}$, which is equivalent to a reduction of $20 \cdot \log(\text{118 GHz}/\text{193 THz GHz}) \approx -64$ dB in phase noise power spectral density \cite{Liu2024}. This estimate assumes perfect correlation of thermal noise in the two cavity modes, and therefore neglects the slight variations in cavity mode volume with wavelength. But for two nearby modes, this approximation sufficiently justifies how the phase noise of the mm-wave beat can surpass the cavity thermal limit. It is then limited by the PDH locking noise of the lasers, which can be addressed through technical advances. 

%The correlations in $\sigma_{th}$ are directly connected to fractional fluctuations in the cavity length $\sigma_L/L$ which reduces (divides) the fluctuations of the FSR as measured by the two PDH locked lasers. We can express this as $\frac{\delta L}{L} = \frac{\text{N}\cdot\delta\text{FSR}}{\text{N}\cdot \text{FSR}}$. 

In the experiment, the two PDH-locked lasers are mixed with a commercial W-band photodetector and the resulting output is a millimeter-wave ("output" in Fig.\ref{1}), having power of $\approx 1 \hspace{3pt}\text{mW}$. Here, we demonstrate millimeter-wave signals at $4\times \text{FSR}$ and $5\times \text{FSR}$, with $94.5 \hspace{3 pt}\text{GHz}$ and $118.1 \hspace{3 pt}\text{GHz}$ carriers respectively. The phase noise of the millimeter-wave signals was measured using cross correlation techniques to remove mixer and amplifier noise.  In brief, the photodetected signal was split with a -3 dB waveguide coupler and each copy was mixed down to the RF using harmonic mixers driven by independent commercial synthesizers. The two resulting baseband signals near $5 \hspace{3 pt}\text{MHz}$ where then cross-correlated on a phase noise analyzer.

\section*{Results}

\begin{figure}[!b]
  \centering
  \includegraphics[width = \linewidth]{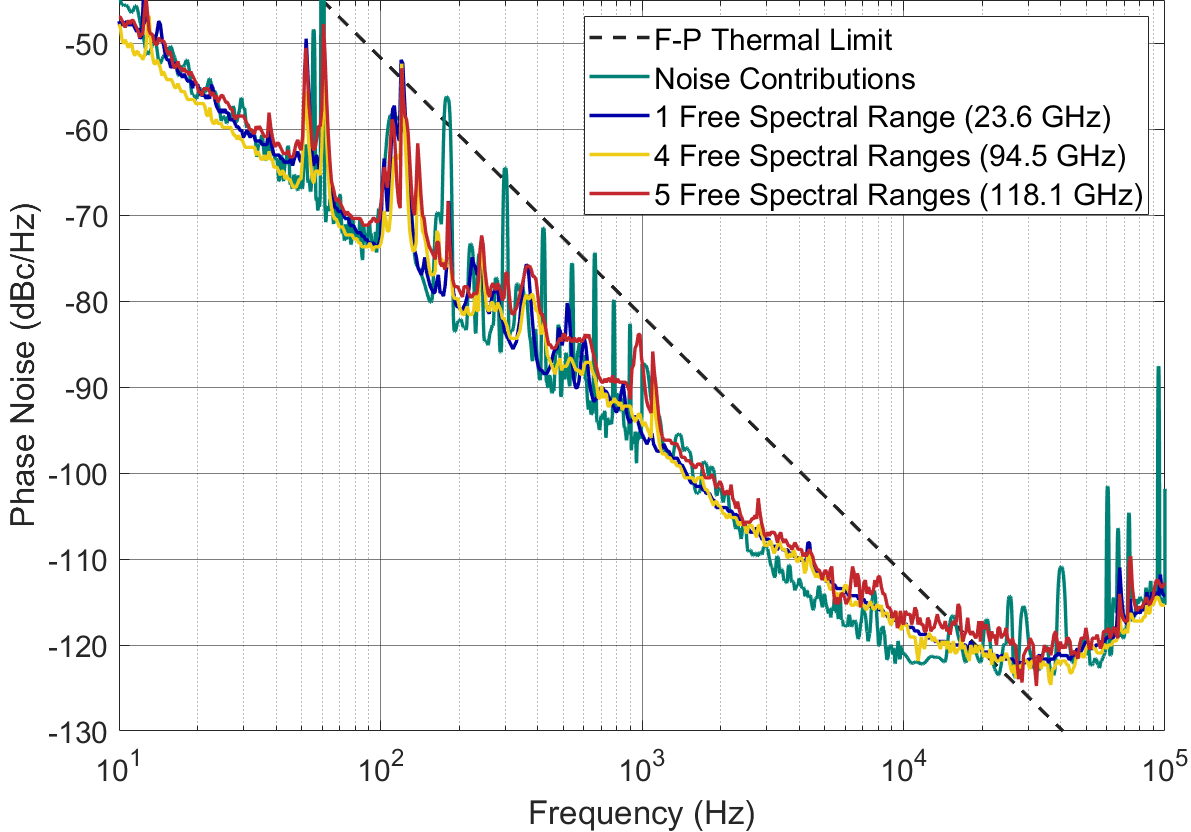}
\caption{Phase noise plot of the 1 FSR (23.6 GHz, Blue), 4 FSR (94.5 GHz, Yellow), and 5 FSR (118.1 GHz, Red) signals generated from heterodyning two self-injection-locked distributed feedback lasers after PDH locking to two modes of the miniature F-P. In green, the phase noise of the cumulative noise contributions is plotted, extrapolated from the PDH-locked laser in-loop noise.}
 \label{2}
 \vspace{-.32 cm}
\end{figure}

The result of the cross-correlation phase noise measurements of the millimeter-wave signals is plotted in Fig. \ref{2}. We see three overlaid traces of approximately the same phase-noise level as the 1 FSR (23.6 GHz, blue), 4 FSR (94.5 GHz, yellow), and 5 FSR (118.1 GHz, red) signals. This indicates that the phase noise of any millimeter-wave signal of integer-multiple FSR frequency can be generated and will inherit the phase noise of the PDH-locked SIL lasers, with no loss in fidelity. The only limiting factor then is the power on the high-speed photodetector, and hence the signal-to-noise ratio, of the heterodyned signals as the frequency difference between the two lasers is increased. The discrepancies in the phase noise of the 118.1 GHz signal versus the 23.6 GHz signal (near 10 kHz offset) arise from the requirement of longer averaging times due to the noise constraints of the cross-correlation reference signals. 

Lastly and for comparison, on the same plot we have plotted (in green) the cumulative uncorrelated noise contributions of various sources, which have been added in quadrature. These sources include the in-loop phase noise of each individual SIL laser during PDH-lock operation, and the SNR floor of the PDH servo loop. The quadrature sum shows good agreement with the measured 1 FSR, 4 FSR, and 5 FSR phase noise. Consequently, lower phase noise lasers or higher PDH-lock gain would offer even greater noise reduction.

\begin{figure}[!b]
\centering\includegraphics[width=\linewidth]{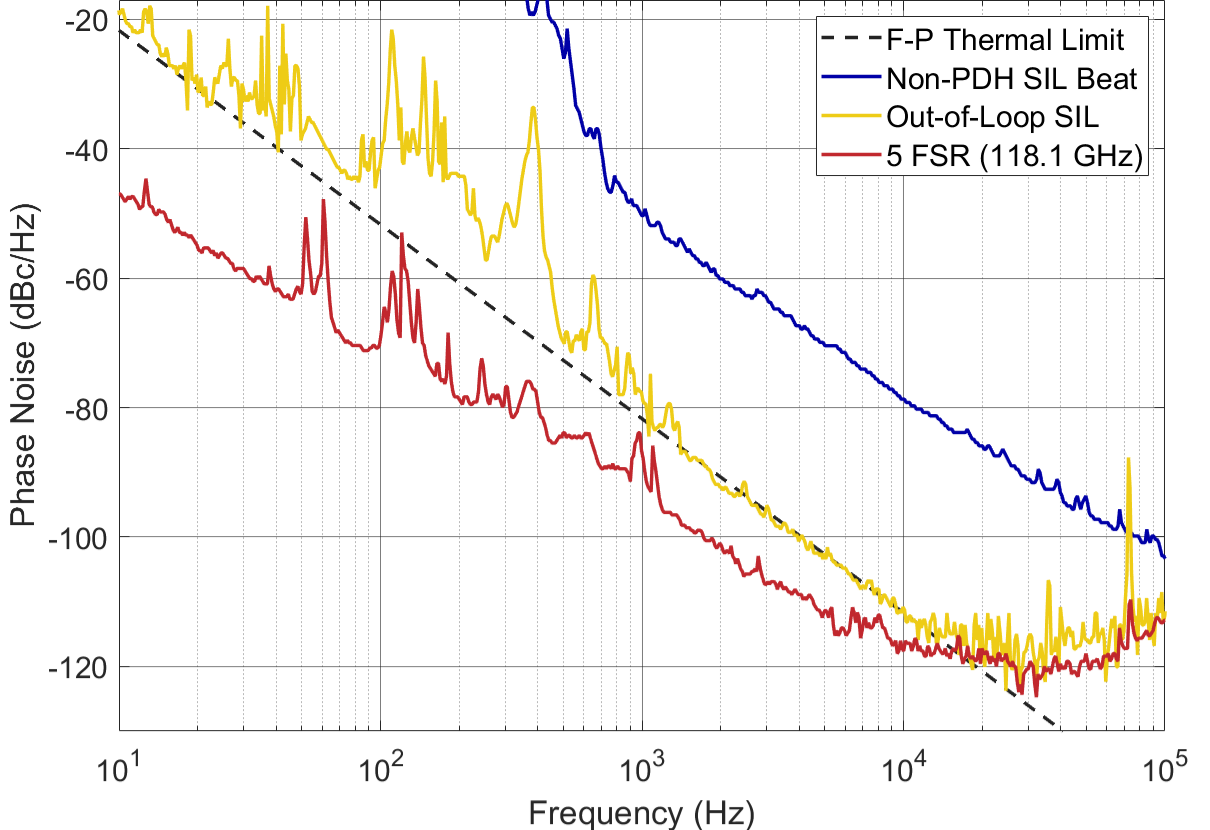}
\caption{(a). Phase noise plot of the PDH-locked (Red) versus unlocked (Blue) SIL lasers. In black-dashed line, the F-P cavity thermal limit referenced from \cite{kelleher2023compact}.}
 \label{3}
 \vspace{-.32 cm}
\end{figure}

In Fig. \ref{3}, we have plotted the free-running (without PDH locking) SIL laser phase noise (blue) and the $118.1$ GHz signal phase noise (red), side-by-side, showing that $\approx 37$ dB of noise reduction at 10 kHz offset is achieved by PDH locking, with AOM and EOM feedback, to the F-P cavity. Here we measured the free-running SIL noise by heterodyning the two SIL lasers on an MUTC $\approx 10 \hspace{3 pt}\text{GHz}$ apart from one another. The $10 \hspace{3 pt}\text{GHz}$ signal is mixed down to DC and measured against a maser reference on a phase noise analyzer. We make the assumption that both SIL lasers have similar noise profiles, such that this phase noise is simply twice (+3 dB) that of a single SIL laser. 

We also plotted the calculated thermal noise floor of the F-P cavity in dashed-black \cite{kelleher2023compact}. Comparing the thermal limit to the 118.1 GHz millimeter-wave signal reveals that for offset frequencies below 10 kHz, the phase noise of the millimeter-wave signal surpasses the thermal-limited optical phase noise over a wide range of Fourier frequencies. 
To further quantify the amount of common-mode rejection exhibited, we independently measured the absolute (out-of-loop) phase noise of an individual PDH-locked SIL laser (yellow in Fig. \ref{3}). As expected, we see that SIL laser noise follows the thermal limit of the F-P cavity up to about 10 kHz. At higher frequencies, the laser noise is limited by PDH servo gain and bandwidth.

\par
The difference between the out-of-loop optical phase noise and the mm-wave phase noise is the amount of noise common to both lasers and hence subtracted (rejected) in the heterodyne signal. Thus the mm-wave signal here gains 22 dB of noise rejection at 10 Hz offset, due to both optical references being PDH locked to the same cavity.

\begin{figure}[!htb]
  \centering
  \includegraphics[width = \linewidth]{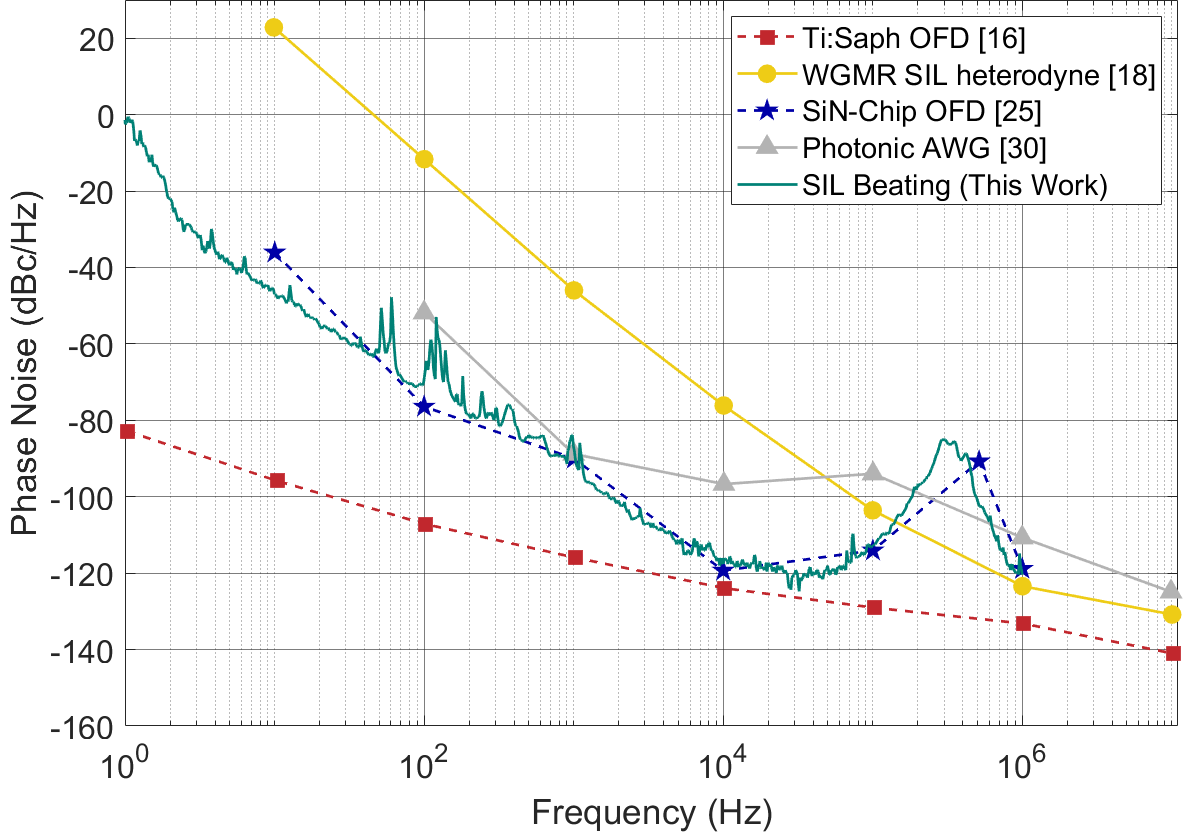}
\caption{Phase noise plot of the millimeter-wave generation landscape scaled to 118.1 GHz \cite{Fortier2016, sun2024integrated, Kittlaus2021,Song2008}.}
 \label{4}
 \vspace{-.32 cm}
\end{figure}

To put our result into context, we have included a survey of the reported phase noise of state-of-the-art photonic millimeter-wave synthesis. As seen in Fig.~\ref{4}, our SIL and PDH framework allows for exceptionally low-noise millimeter-wave synthesis, largely falling well below commercial products and existing photonic-based sources generation up until the servo bump at around 300 kHz. This does not tell the full story, however, as other millimeter-wave generators rely on significantly more complex systems, to achieve lower and similar phase noise.

For example, the lowest phase noise achieved and plotted in red, incorporates full optical frequency division using an octave spanning, titanium-doped sapphire, bulk-optic frequency comb and a filtering etalon \cite{Fortier2016}. The power required to run such a system is thus significantly higher than the one in this work. The lowest photonic-chip-based mm-wave noise (blue), falls directly in line with this work's, yet requires a microcomb to achieve OFD, which inevitably requires a more complex locking scheme, while simultaneously requiring two lasers PDH locking to a microresonator cavity \cite{sun2024integrated}. In this work, we've shown that these phase-noise levels can be achieved with a simple, passive self-injection locking and PDH scheme, requiring low input currents (200 mA to drive the lasers) and providing a path toward chip integration.

The phase noise survey also includes research mm-wave synthesizers (grey) \cite{Song2008} and a similar self-injection-locking heterodyne scheme (yellow) but without PDH locking to a cavity. 

\section*{Conclusion and Outlook} 

In conclusion, we have shown a simplified, compact way to generate low phase-noise millimeter-wave carriers, by transferring the noise of stable optical references to any integer multiple of the F-P cavity FSR.  Importantly, through common-mode rejection, we  surpass the thermal limit of the reference cavity. In doing so, we have shown millimeter-wave phase noise of $-117 \hspace{3pt}\text{dBc/Hz}$ at 10 kHz offset, regardless of carrier frequency. This performance is better than any chip-scale photonic mm-wave generation to date, and is within 5 dB the highest performing full-OFD millimeter-wave source. 

In addition to the phase noise performance, the low-complexity and compact form factor of this experiment both lend themselves towards the argument of full integratability.  For example, with advances of air-gap micro- F-P cavities \cite{Jin2022, Liu2024, Cheng2023}, integrated SIL lasers and on-chip MUTC photodetectors \cite{Xiang2023, Zang2018reduction}, the approach we present can be integrated onto a single chip. And in comparison to more complex OFD schemes \cite{Kudelin2024, sun2024integrated}, we show an attractive option to reduce the number of integrated components and realize on-chip, low-noise mm-wave generation with robust and turn-key operation. 

%On a related note, heterogeneously integrated lasers and microresonators would allow for more stable self-injection locking, allowing for more stable PDH-locked references, hence lower phase-noise mm-waves. With both SIL references sharing the same chip, it is possible that some level of common noise will be shared between them, which would be cancelled out in the beat note. This would have important implications for a scheme in which one passively locks the two lasers to such a cavity; the servo bump would be remediated and the carrier signal could follow the cavity thermal noise trend ($1/f^3$) to the shot noise limit. 

\bibliographystyle{ieeetr}
\bibliography{SILBeatMMBib}

\medskip
\begin{footnotesize}

\noindent \textbf{Corresponding authors}: \href{mailto:william.groman@colorado.edu}{william.groman@colorado.edu} and \href{mailto:scott.diddams@colorado.edu}{scott.diddams@colorado.edu}

\noindent \textbf{Funding}: 
This research was supported by DARPA GRYPHON program (HR0011-22-2-0009) and NIST.% National Aeronautics and Space Administration
(80NM0018D0004).

\noindent \textbf{Acknowledgments}: 
Commercial equipment and trade names are identified for scientific clarity only and does not represent an endorsement by NIST.

\noindent \textbf{Author Contributions}:

W.G., I.K., J.E.B., F.Q., and S.A.D. conceived the experiment and supervised the project.
W.G., I.K., and S.A.D. wrote the paper with input from all authors.
W.G., I.K., and A.L.  built the experiment and performed the mm-wave experiment.
D.L., T.N., and Y.L. provided mm-wave reference sources and input regarding the cavity and cross-correlation measurements. 
M.L.K and C.A.M. built the cavity and gave crucial information regarding the cavity. 
J.G., L.W., W.J., and J.E.B provided the lasers and spiral microresonators, as well as input regarding the operation of the lasers. 
%L.W., prepared the DFB laser butterfly packages for the experiment.
%Q.-X.J., J.G., W.J., L.W., and C.X. prepared the microcomb and spiral resonators for the experiment.
%M.L.K., and F.Q. built the F-P cavity
%D.L., T.N., C.A.M., Y.L., and F.Q. provided optically derived microwave reference and aided in the microwave phase noise measurement system.
%J.B. and J.C.C. provided MUTC detectors.
All authors contributed to discussion of the results.

\noindent \textbf{Disclosures}: 

The authors declare no conflicts of interest.

\noindent \textbf{Data and materials availability}:  

All data for the figures in this manuscript is available at \url{}.

\end{footnotesize}

\end{document}